\documentclass[aps,prd,showpacs,twocolumn,superscriptaddress,nofootinbib,preprintnumbers]{revtex4-1} 

\newcommand{\be}{\begin{equation}}
\newcommand{\ee}{\end{equation}}
\newcommand{\beqa}{\begin{eqnarray}}
\newcommand{\eeqa}{\end{eqnarray}}

\makeatletter
\def\p@subsection{}
\makeatother

\usepackage{color,graphicx,amsmath,amssymb,lipsum,bm}
\usepackage{graphicx}% Include figure files
\usepackage{dcolumn}% Align table columns on decimal point
\usepackage{bm}% bold math

\usepackage{hyperref}
\hypersetup{colorlinks=true,citecolor=blue,linkcolor=blue,urlcolor=blue, backref=false,pdfborder={0 0 0}}

\usepackage[normalem]{ulem}
\usepackage[utf8]{inputenc} 
\usepackage{mathtools}
 
\usepackage{float}
\usepackage{aas_macros}
\usepackage{multirow}

\usepackage{ulem}
\usepackage{diagbox}

\usepackage[dvipsnames]{xcolor}
\usepackage{mathrsfs}

\definecolor{xlinkcolor}{rgb}{0.7752941176470588, 0.22078431372549023, 0.2262745098039215}

\usepackage{float}

\begin{document}

\preprint{MIT-CTP/5954}

\title{A-BHPT-toolkit: 
Analytic 
Black Hole Perturbation Theory Package \\ for Gravitational Scattering Amplitudes }

\author{Jovan Markovic}
\email{jm2725@cam.ac.uk}
\affiliation{Massachusetts Institute of Technology, 
Cambridge, MA 02139, USA}
\affiliation{University of Cambridge, Cambridge CB3 0HE, United Kingdom}

\author{Mikhail M. Ivanov }
\email{ivanov99@mit.edu}
\affiliation{Center for Theoretical Physics -- a Leinweber Institute, Massachusetts Institute of Technology, 
Cambridge, MA 02139, USA} 
 % \affiliation{The NSF AI Institute for Artificial Intelligence and Fundamental Interactions, Cambridge, MA 02139, USA}

\begin{abstract}
Applications of effective field theory (EFT) and scattering amplitudes 
to gravitational problems have recently
produced many unique results that 
advanced our understanding of 
the dynamics of compact binaries. 
Many of these results were made possible 
by comparing gravitational scattering amplitudes in EFT 
with exact expressions from general relativity. However, the latter expressions
are not easily available as they require intricate solution techniques 
for the Teukolsky master equation, such as
the Mano-Suzuki-Takasugi (MST) method.
In this paper, we develop and present the first public package that enables computations of 
gravitational scattering amplitudes
in black hole perturbation theory 
within the MST framework. 
Our package supports both analytic computations 
to a given post-Minkowskian (PM) order in the low-frequency limit and exact numerical
computations for an arbitrary 
frequency of the 
perturbing field. 
As an application, we compute 
scattering 
phase shifts
and inelasticity parameters
for massless spin -- 0, 1, and 2 fields
resulting from scattering off a rotating Kerr black hole 
through the third PM order
and compare these results with the exact
numerical solutions.
Our package is publicly available
at \url{https://github.com/Jovan888777/bh-scattering-phase/tree/main}.
\end{abstract}

\maketitle

\section{Introduction}

Progress in gravitational wave
astronomy~\cite{LIGOScientific:2014pky,VIRGO:2014yos,LIGOScientific:2016aoc,LIGOScientific:2018mvr,LIGOScientific:2020ibl,LIGOScientific:2021usb,LIGOScientific:2021djp,KAGRA:2020agh}, culminating with the most
recent release of the first part of the LIGO-Virgo-KAGRA data
from the fourth observing run (O4a)~\cite{LIGOScientific:2025snk},
has provided strong motivation 
for novel theoretical 
developments in gravitational physics. 
In particular, the application 
of techniques from the high-energy physics
toolbox, such as effective field theories
(EFTs) 
and scattering amplitudes~\cite{Goldberger:2004jt,
Goldberger:2005cd,Porto:2007qi,Rothstein:2014sra,Goldberger:2022rqf,Goldberger:2022ebt,Porto:2016pyg,Kalin:2020mvi,Mogull:2020sak,Cheung:2018wkq,Kosower:2018adc,Bern:2019crd,Bern:2020buy,Buonanno:2022pgc,Cheung:2023lnj,Kosmopoulos:2023bwc,Bjerrum-Bohr:2022blt,Bautista:2023sdf,Brandhuber:2023hhy,Herderschee:2023fxh,Elkhidir:2023dco,Georgoudis:2023lgf,Caron-Huot:2023vxl}, 
have produced many new results that 
have significantly advanced 
our understanding of gravitational dynamics. 
For example, gravitational 
scattering amplitudes
can be used to define and extract 
tidal responses of black holes
and exotic compact objects
in a manifestly gauge-invariant
and coordinate-independent fashion. 
For instance, the conservative 
tidal Love numbers that capture
compact 
body's responses to external gravitational
fields are defined as Wilson 
coefficients in the worldline 
EFT~\cite{Goldberger:2004jt,Goldberger:2007hy}.\footnote{See~\cite{Charalambous:2021mea,Charalambous:2021kcz,Charalambous:2022rre,Charalambous:2023jgq,Charalambous:2025ekl,Hui:2020xxx,Hui:2021vcv,Hui:2022vbh,Ivanov:2022hlo,Ivanov:2023hlo} for 
detailed
discussions about Love numbers of black holes and their various definitions and extraction modes, as well as relations to hidden symmetries.}
The universality of EFT then allows one to match Love numbers 
from a simple scattering amplitude
calculation
and then use the results of this measurement
to make predictions for gravitational waveforms~\cite{Goldberger:2007hy,Ivanov:2022qqt,Saketh:2023bul,Ivanov:2024sds,Ivanov:2025ozg}. 

From the practical perspective, 
gravitational scattering amplitudes
in EFT have to be matched 
to scattering amplitudes in general 
relativity in order to extract 
these 
Wilson coefficients. 
In the context of black holes, 
the 
execution of this program requires
one to have analytic expressions
for scattering amplitudes, which 
in principle can be computed
from black hole perturbation theory 
(BHPT)~\cite{matzner1968scattering,1978ApJS...36..451M,Matzner:1977dn}.
based on the Teukolsky master equation~\cite{Teukolsky:1972my,Teukolsky:1973ha,teukolsky1974perturbations}.
In particular, BHPT equations
admit an exact solution 
in terms of a series over 
hypergeometric functions 
via the Mano-Suzuki-Takasugi (MST) method~\cite{Mano:1996gn,Mano:1996mf,Mano:1996vt,Sasaki:2003xr}.\footnote{See also 
\cite{Bonelli:2021uvf,Bonelli:2022ten,Bautista:2023sdf,Castro:2013kea,Castro:2013lba,Nasipak:2024icb} for alternative techniques.}
In practice, however, 
the analytic scattering amplitude
calculations in BHPT are quite contrived 
and even though they have been computed 
in a number of papers, e.g.~\cite{Dolan:2008kf,Bautista:2022wjf,Ivanov:2022qqt,Saketh:2023bul,Ivanov:2024sds,Caron-Huot:2025tlq}, 
it is not easy to obtain them 
from scratch. This contrasts with the 
situation in other aspects of BHPT, 
where excellent numerical tools, such as
\texttt{Black Hole Perturbation Toolkit} package~\cite{BHPToolkit}, are readily 
available.\footnote{~\url{bhptoolkit.org}} In this work, we fill this gap 
and provide a first public package 
for analytic 
computations of BHPT scattering 
amplitudes. 

Our paper is structured as follows. 
We first critically review 
analytic calculations on BHPT scattering 
amplitudes 
from the Teukolsky master equation,
and clearly write down all the ingredients
needed for such calculations in Section~\ref{sec:background}. 
Then we present a method of building 
the analytic solution
to a given post-Minkowskian 
order (order in Newton's gravitational constant $G$)
using the MST method in Section~\ref{sec:res}. 
There we also present the results of our method, which includes 
several novel results, such as  
expressions for scattering phase
shifts
of the gravitational Raman process
with a non-zero black hole spin. 
Section~\ref{sec:concl} draws conclusions. Appendices A and B
contain additional tables and plots.

\section{Background}
\label{sec:background}

\subsection{Scattering in BHPT from the Teukolsky Master Equation}
% Solving the underlying field equations directly in full to describe scattering is difficult because the equations are non-linear and coupled, requiring use of numerical simulation to obtain solutions. 
We seek a perturbative solution
to scattering amplitudes 
in black hole perturbation theory using the Teukolsky master equation~\cite{Teukolsky:1972my}, based on the Newman-Penrose formalism~\cite{newman1962approach,newman1963errata}. In the following discussion, $M$ denotes the mass of the black hole, $a$ its spin parameter, and $(r, \theta, \phi)$ the spherical coordinates centered at the black hole singularity.

We consider three types of massless fields: the spin-0 scalar field, the spin-1 electromagnetic field, and the spin-2 perturbation of the gravitational field described by the Teukolsky equation. The field perturbation are captured by 
the Neumann-Penrose scalars $\Psi_i$, which separate into radial and angular components. For example, for $\Psi_4$, we may write
\begin{equation}
    \Psi_4 \propto e^{-i\omega t}e^{im\phi} {}_{s}S^m_l(\theta; a\omega) {}_{s}R_{lm\omega}(r),
    \label{eq:Neu_Pen}
\end{equation}
which causes the perturbative Newman-Penrose equations to separate into
\begin{subequations}
    \begin{multline}
    \label{eq:Teuk_rad}
        \Delta^{-s}\frac{d}{dr}\left(\Delta^{s + 1}\frac{dR}{dr}\right) + \Big(\frac{K^2 - 2is(r - M)K}{\Delta} \\
        + 4is\omega r - \lambda_{lm}\Big)R = 0,
    \end{multline}
    \begin{multline}
    \label{eq:Teuk_ang}
        \frac{1}{\mathrm{sin}\ \theta}\frac{d}{d\theta}\left(\mathrm{sin}\ \theta \frac{dS}{d\theta}\right) + \Big(a^2\omega^2\mathrm{cos}^2\theta - \frac{m^2}{\mathrm{sin}^2 \theta}
        - \frac{2ms\ \mathrm{cos}\ \theta}{\mathrm{sin}^2\theta} \\ 
        - 2a\omega s \ \mathrm{cos}\ \theta - s^2 \mathrm{cot}^2\theta + s + A_{lm}\Big)S = 0. 
    \end{multline}
\end{subequations}
Here, $\Delta = r^2 - 2Mr + a^2$, $K = (r^2 + a^2)\omega - am$, and $\lambda_{lm} = A_{lm} + a^2\omega^2 - 2am\omega$. We set $s = 0$ for scalar perturbations, $s = 1$ for electromagnetic perturbations, and $s = 2$ for gravitational perturbations. For a fixed $s$, the solution space has a basis enumerated by the integers $l$ and $m$, for which it holds $|s| \leq l$ and $l \geq m \geq -l$, as well as by the nonnegative real angular frequency $\omega$.  

 To proceed, we seek a solution as an expansion in the parameter $\epsilon = 2M\omega$ as is standard 
 in the MST approach. 
The angular equation~\eqref{eq:Teuk_ang} has well known solutions through the spheroidal harmonics. These, as well as the value of the eigenvalue $\lambda_{lm}$ have known expansions in $\epsilon$ to arbitrary order \cite{Berti_2006}, and are provided through the Black Hole Perturbation Toolkit (BHPT toolkit) \cite{Black_Hole_Perturbation_Toolkit}.

Now let us discuss 
the radial equation~\eqref{eq:Teuk_rad}. To start, define the tortoise coordinate $r^*$ to satisfy $dr_*/{dr}=(r^2 + a^2)/\Delta$, the horizons of the rotating black hole as $r_{\pm} = M \pm \sqrt{M^2 - a^2}$, and the quantity $\Omega_h = a/(2Mr_+)$. Then, considering extremal solutions to~\eqref{eq:Teuk_rad}, we recover
\begin{widetext}
\begin{equation}
    {}_{s}R_{lm\omega}(r) \sim
    \begin{cases}
        B^{(trans)}_{lm\omega}\Delta^{-s}e^{-i(\omega - m\Omega_h)r_*} & \text{for } r_* \rightarrow -\infty \\
        B^{(inc)}_{lm\omega}r^{-1}e^{-i\omega r_*} + B^{(refl)}_{lm\omega}r^{-2s - 1} e^{+i\omega r_*} & \text{for } r_* \rightarrow +\infty.
    \end{cases}
    \label{eq:rad_extrem}
\end{equation}
\end{widetext}
These relations set the asymptotic 
scattering 
boundary conditions 
at infinity. 

\subsection{Scattering and Absorption}
We are particularly interested in a characteristic phase factor related to the exponentials in~\eqref{eq:rad_extrem} as $r^* \rightarrow +\infty$, defined as
\begin{equation}
\label{eq:fulldelta}
    e^{2i_s\delta_{lm\omega}^P} = (-1)^{l + 1} \frac{_{s}C^m_{l}(a\omega)}{(2\omega)^{2s}}\frac{_{-s}B^{(refl)}_{lm\omega}}{_{-s}B^{(inc)}_{lm\omega}}.
\end{equation}
Here, $_{s}C_l^m(a\omega)$ is the Teukolsky-Starobinsky constant \cite{Casals_2021}, which has values of
\begin{widetext}
\be
\begin{split}
  & {}_0C_l^m(a\omega) = 1, \\
 & {}_1C_l^m(a\omega) = \Big[ \big({}_1\lambda_{lm} + 2 + a^2\omega^2 - 2am\omega\big)^2 
 + 4am\omega - 4a^2\omega^2 \Big]^{1/2}\,,\\
 &
  {}_2C_l^m(a\omega) = \Big[ \big[ \big(_{-2}\lambda_{lm} + 2\big)^2 + 4am\omega - 4a^2\omega^2 \big]
\times \big[ \big(_{-2}\lambda_{lm}\big)^2 + 36am\omega - 36a^2\omega^2 \big]\\
& + \big(2\,{}_{-2}\lambda_{lm} + 3\big)(96a^2\omega^2 - 48am\omega) 
\Big]^{1/2} + 12iPM\omega.
\end{split}
\ee
\end{widetext}
For $s = 2$ only, the variable $P = \pm 1$ denotes parity.

The phase $e^{2i\delta_{lm\omega}^P}$ can be related to important quantities in scattering theory. For example, as in \cite{Dolan:2008kf}, for $s = 2$ the scattering cross section is given by
\begin{equation}
    \frac{d\sigma}{d\Omega} = |f(\theta)|^2 + |g(\theta)|^2,
\end{equation}
where the partial wave series take the values
\begin{subequations}
    \begin{multline}
        f(\theta) = \frac{\pi}{i\omega} \sum_{P = \pm 1} \sum_{l = 2}^{\infty} {}_{-2}S_{l}^{2}(0;a\omega)\ {}_{-2}S^{2}_{l}(\theta; a\omega) \\
        \cdot [e^{2i\delta_{l2\omega}^P} - 1],
    \end{multline}
    \begin{multline}
        g(\theta) = \frac{\pi}{i\omega}\sum_{P = \pm 1} \sum_{l = 2}^{\infty} P(-1)^l _{-2}S_{l}^2(0;a\omega) _{-2}S^2_l(\pi - \theta; a\omega) \\
        \cdot [e^{2i\delta_{l2\omega}^P} - 1].
    \end{multline}
\end{subequations}
The absorption cross section is determined by
\begin{equation}
    \sigma_a = \frac{4\pi^2}{\omega^2} \sum_{l = 2}^{\infty}|_{-2}S^2_l(0; a\omega)|^2 \mathbb{T}_{l2},
\end{equation}
with the transmission factor $0 \leq \mathbb{T}_{lm} \leq 1$ given by
\begin{equation}
\label{eq:trans}
    \mathbb{T}_{lm} = 1 - |e^{2i\delta^P_{lm\omega}}|^2.
\end{equation}
Similarly, one can relate these cross sections to the imaginary parts of  $\delta_{lm\omega}$ for $s = 0, 1$. Therefore, determining the phase factor is of critical importance for describing scattering around rotating black holes.

\subsection{The Mano-Suzuki-Takasugi Method}
To proceed, the ratio $_{-s}B^{(refl)}_{lm\omega}/_{-s}B^{(inc)}_{lm\omega}$ needs to be determined and inserted into eq.~\eqref{eq:fulldelta}. As discussed, we seek an expansion of the solution in low $\epsilon = 2M\omega$. To that end, we invoke the Mano-Suzuki-Takasugi (MST) formalism \cite{Mano:1996vt}.

Following it, we obtain a series expansion in terms of hypergeometric functions for $_sR_{lm\omega}(r)$, which converges everywhere except for $r \rightarrow \infty$. Similarly, we obtain a solution in terms of Coulomb wave functions that converges everywhere outside the event horizon. Matching these two solutions at the horizon, we obtain the full solution. To tackle scattering, we are particularly interested in the hypergeometric expansion, which converges near the singularity. The ingoing solution to eq.~\eqref{eq:Teuk_rad} can be written
\begin{equation}
\label{eq:radsol}
    R_{in}^{\nu} = e^{i\epsilon\kappa x}(-x)^{-s - i\frac{\epsilon + \tau}{2}}(1 - x)^{i\frac{\epsilon - \tau}{2}}p_{in}^{\nu}(x),
\end{equation}
where we have defined $q = a/M$, $\kappa = \sqrt{1 - q^2}$, $\tau = (\epsilon - mq)/\kappa$, and $x = \omega(r_+ - r)/(\epsilon\kappa)$. For $p_{in}^{\nu}(x)$, we have the expansion
\begin{equation}
    p_{in}^{\nu}(x) = \sum_{n = -\infty}^{\infty}{a_n^\nu p_{n + \nu}(x)},
\end{equation}
where $p_{n + \nu}(x)$ denotes the standard hypergeometric function
\begin{equation}
p_{n + \nu}(x) = F(n + \nu + 1 - i\tau, -n - \nu - i\tau; 1 - s - i\epsilon - i\tau; x).
\end{equation}
The parameter $\nu$ in the above equation
is called the renormalized angular momentum, which generalized
the orbital quantum number $l$
beyond the Newtonian limit. 
We discuss its determination shortly. 

After inserting~\eqref{eq:radsol} into eq.~\eqref{eq:Teuk_rad}, we get that the expansion coefficients $a_n^{\nu}$ satisfy the three-term recurrence relation
\begin{equation}
\label{eq:3-point}
    \alpha_n^{\nu} a_{n + 1}^\nu + \beta_n^{\nu}a_n^{\nu} + \gamma_n^{\nu}a_{n - 1}^\nu = 0,
\end{equation}
with the recurrence coefficients given by
\begin{subequations}
    \begin{multline}
        \alpha_n^{\nu} = 
        \frac{i\epsilon\kappa(n + \nu + 1 + s + i\epsilon)}{n + \nu + 1} \\
        \cdot \frac{(n + \nu + 1 + s - i\epsilon)(n + \nu + 1 + i\tau)}{2n + 2\nu + 3},
    \end{multline}
    \begin{multline}
        \beta_n^{\nu} = -\lambda_{lm} - s(s + 1) + (n + \nu)(n + \nu + 1) + \epsilon^2 \\
        + \epsilon(\epsilon - mq) + \frac{\epsilon(\epsilon - mq)(s^2 + \epsilon^2)}{(n + \nu)(n + \nu + 1)},
    \end{multline}
    \begin{multline}
        \gamma_n^{\nu} = - \frac{i\epsilon\kappa(n + \nu - s + i\epsilon)(n + \nu - s - i\epsilon)(n + \nu - i\tau)}{(n + \nu)(2n + 2\nu - 1)}.
    \end{multline}
\end{subequations}
Conventionally, we take $a_0^\nu = 1$. An important point to note is that $\nu$ is a new parameter. To fix it, the solutions of the three-point relation \eqref{eq:3-point} emerging from $n \rightarrow -\infty$  match those of $n \rightarrow \infty$. This procedure is carefully examined in the next section.

To determine $\nu$, raising and lowering continued fractions for MST coefficients are first defined as
\begin{subequations}
    \begin{equation}
    \label{eq:raising}
        R_n(\nu) = \frac{a_n^{\nu}}{a_{n - 1}^{\nu}} = -\frac{\gamma_n^{\nu}}{\beta_n^{\nu} + \alpha_n^{\nu}R_{n + 1}(\nu)},
    \end{equation}
    \begin{equation}
    \label{eq:lowering}
        L_n(\nu) = \frac{a_n^{\nu}}{a_{n + 1}^{\nu}} = -\frac{\alpha_n^{\nu}}{\beta_n^{\nu} + \gamma_n^{\nu}L_{n - 1}(\nu)}.
    \end{equation}
\end{subequations}
In the expansion of the MST recurrence relation in $\epsilon$, the minimal solution for $n \rightarrow \infty$ is matched with the solution for $n \rightarrow -\infty$ by imposing connection formulas, an alternate form of equation~\eqref{eq:3-point}, between the raising and lowering ratios
\begin{equation}
\label{eq:con_form}
    \beta_u^{\nu} + \alpha_u^{\nu}R_{u + 1}(\nu) + \gamma_u^{\nu}L_{u - 1}(\nu) = 0.
\end{equation}

Now, after expansion solutions to~\eqref{eq:Teuk_rad} are obtained, we may determine $_{-s}B^{(refl)}_{lm\omega}/_{-s}B^{(inc)}_{lm\omega}$. First, we write
\begin{subequations}
    \begin{multline}
    \label{eq:Aplus}
        A_{+}^{\nu} = e^{-(\pi/2)\epsilon}e^{(\pi/2)i(\nu + 1 - s)}2^{-1 + s - i\epsilon} \\
        \cdot \frac{\Gamma(\nu + 1 - s + i\epsilon)}{\Gamma(\nu + 1 + s - i\epsilon)}\sum_{n = -\infty}^{\infty}a_n^{\nu},
    \end{multline}
    \begin{multline}
    \label{eq:Aminus}
        A_{-}^{\nu} = e^{-(\pi/2)\epsilon}e^{-(\pi/2)i(\nu + 1 + s)}2^{-1 - s + i\epsilon} \\
        \cdot \sum_{n = -\infty}^{\infty} (-1)^n \frac{(\nu + 1 + s - i\epsilon)_n}{(\nu + 1 - s + i\epsilon)_n}a_n^{\nu}.
    \end{multline}
\end{subequations}
Finally, denoting $K_\nu$ the connection coefficient between the hypergeometric and Coulomb solution \cite{Bautista:2022wjf}, we have
\begin{subequations}
\begin{multline}
\label{eq:Binc}
    _sB_{lm\omega}^{(inc)} = \omega^{-1}\Big(K_\nu - ie^{-i\pi\nu}\frac{\sin(\pi(\nu - s + i\epsilon)}{\sin(\pi(\nu + s - i\epsilon)}\Big) \\
    \cdot e^{-i\big(\epsilon \ln \epsilon - \frac{1 - \kappa}{2}\epsilon \big)}A_{+}^{\nu},
\end{multline}
\begin{multline}
\label{eq:Brefl}
    _sB_{lm\omega}^{(refl)} = \omega^{-1 - 2s}(K_{\nu} + ie^{i\pi\nu}K_{-\nu - 1}) \\ 
    \cdot e^{i\big(\epsilon \ln \epsilon - \frac{1 - \kappa}{2}\epsilon \big)}A_{-}^{\nu}.
\end{multline}
\end{subequations}
Relations~\eqref{eq:Binc} and~\eqref{eq:Brefl} were obtained by considering the behavior of the solution~\eqref{eq:radsol} in appropriate extremal settings of the coordinate $r_*$.

A final point of note is that though the MST solution, when expanded in $\epsilon$, converges well in the radial coordinate, it has a limited radius of convergence in $\epsilon$ \cite{Fujita:2005kng}
if the connection constraint
\eqref{eq:con_form} 
is solved perturbatively in $\epsilon$.
In other words, 
while the MST solution formally 
converges for any frequency
in the numerical sense, the 
analytic solution
has a finite radius of convergence
in $\epsilon$.
For example, for $s = 2, l = 2, m = 0$, a discrepancy appears in the renormalized angular momentum a little above $\epsilon = 0.7$. This is related to the 
development of the imaginary part
for the renormalized angular 
momentum for $\epsilon$ greater than this value~\cite{Sasaki:2003xr}. 
This fact will be evident later when our expansion is compared with a numerical solution obtained through the BHPT toolkit.

\section{Method and Results}
\label{sec:res}

We aim to determine the analytical expansion of the scattering phase shift for low $\epsilon$ to a desired order $n_0$ in a robust, efficient, and repeatable way. To do this, an algorithm that automates most of the process was developed in Mathematica. The diagram below outlines the procedure.
\begin{figure}[!hbt]
    \centering
    \includegraphics[width=1\linewidth]{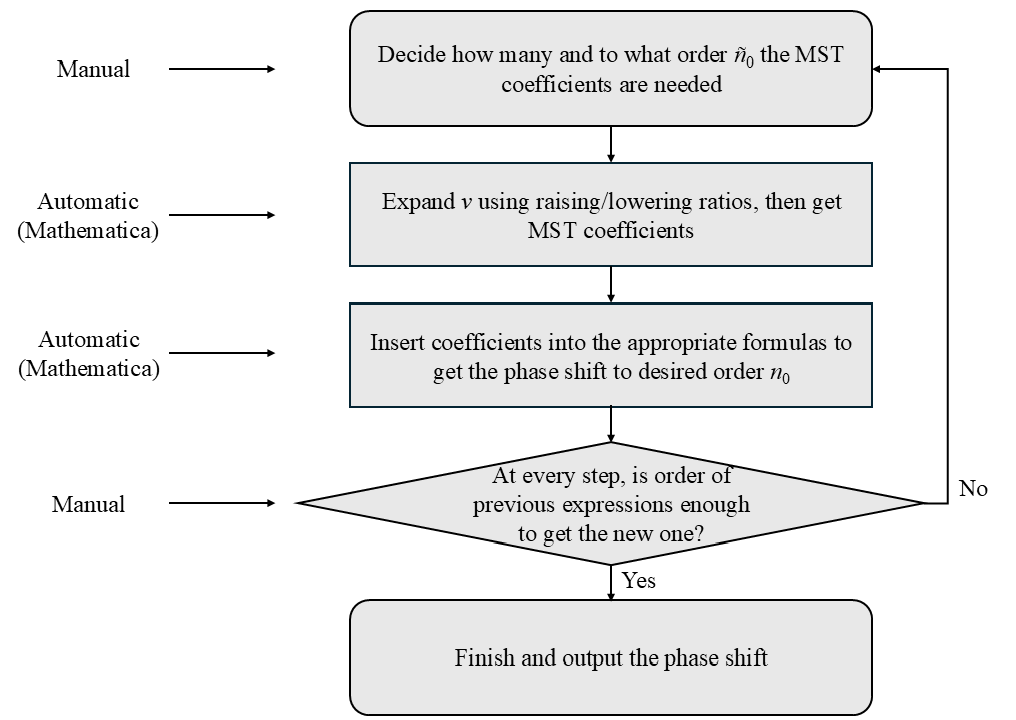}
    \caption{Flowchart of the phase determination procedure. The manual steps are essentially choosing to what order each intermediate expression should be expanded to, while Mathematica carries out the automatic ones.}
    \label{fig:flowchart}
\end{figure}
A notable detail is that the order $\tilde{n_0}$ the MST coefficients are determined to, as well as orders of all intermediate expressions until the phase shift, might need to be bigger than the desired order of the phase shift itself $n_0$. The reason for this is that intermediate expressions might have other factors whose leading order is $O(1/\epsilon^k)$ for $k \geq 1$. To evaluate each of these expressions to a desired order, we need higher order terms in the MST coefficients and $\nu$. Such considerations form the crux of the manual part of the procedure. Due to the high variability of such leading order terms for various starting parameters ($s, l$, and $m$), careful manual analysis is needed at every step to ensure we are obtaining every expression to the desired order.
% Let us give now
% the description of the algorithm
% to compute the MST 
% coefficients. 

The only systematic and user-friendly 
 currently available 
 public 
routine
 for determining relevant scattering parameters is numerical, through the BHPT toolkit. To recover $\nu$, the toolkit numerically solves the transcendental connection equation~\eqref{eq:con_form} via root-finding, after sufficiently expanding the continued fractions in equation~\eqref{eq:lowering} and~\eqref{eq:raising}. Because it avoids analytic expansion of these fractional forms, this method does not suffer from convergence issues at large $\epsilon$. To provide an equally convenient analytical method, we wrote and polished a new Mathematica notebook containing our algorithm. As we detail in the next section, our algorithm closely parallels the numerical approach of BHPT toolkit, with the key difference that we solve the connection equation~\eqref{eq:con_form} analytically, order by order, rather than numerically.

Besides presenting novel analytical expansions of parameters and a ready-for-use analytical algorithm, we also compare the results with BHPT toolkit's numerical solutions. Specifically, for an example set of $s, l$, and $m$, the expanded $\nu$ is compared with the numerical module \texttt{RenormalizedAngularMomentum} from BHPT toolkit's \texttt{Teukolsky} package. In addition, though the BHPT toolkit offers no function to recover the scattering phase factor directly, another new Mathematica notebook was written that first calculates $\nu$ numerically using the toolkit, then determines high-order MST coefficients explicitly using~\eqref{eq:raising} and~\eqref{eq:lowering}, and finally inserts them into~\eqref{eq:Binc},~\eqref{eq:Brefl}, and~\eqref{eq:fulldelta}. This way, convergence issues are avoided as every intermediate quantity is determined numerically. In spirit, this mirrors the way the toolkit uses MST coefficients to obtain desired quantities. Notably, besides a module for numerically determining the phase factor, the newly written notebook also contains a module to test numerical convergence of specified MST coefficients. The obtained numerical phase factor is then compared with our analytical expansion. In both these comparisons, we expect a discrepancy for large $\epsilon$ due to the aforementioned convergence issues of the analytical expansions.

\subsection{An algorithm for MST coefficients}
Firstly, the expansion of the renormalized angular momentum $\nu$ and the MST coefficients $a_n^{\nu}$ is required to order $\tilde{n_0} \geq n_0$ in $\epsilon$. This process corresponds to steps one and two in Figure \ref{fig:flowchart}.

The appropriate connection formula
\eqref{eq:con_form}
also fixes the value of $\nu$. For convenience, we choose $u = 0$. Thus, to fix $\nu$ up to order $\tilde{n}_0$ in $\epsilon$, $R_1(\nu)$ and $L_{-1}(\nu)$ are required to order $\tilde{n_0}$ as well. To get them, the lowermost $\tilde{n_0}$ raising ratios for $n > 0$ are obtained using the continued formula~\eqref{eq:raising}, with $R_{\tilde{n}_0}(\nu)$ calculated with $R_{\tilde{n}_0 + 1}(\nu)$ set to zero, while the uppermost $\tilde{n}_0 + \Delta n$ lowering ratios for $n < 0$ are obtained using~\eqref{eq:lowering} in the same way. The ratios are $O(\epsilon)$, and thus the $n$-th coefficient is $O(\epsilon^{|n|})$ except in certain special cases for $n < 0$ (which is why we calculate $\tilde{n_0} + \Delta n$ lowering ratios instead of $\tilde{n}_0$, setting $\Delta n$ equal up to 3. See \cite{Mano:1996vt} eq. (5.6) for exact order relationships, and to guide the choice of $\Delta n$). The connection equation~\eqref{eq:con_form} is solved order by order in $\epsilon$, obtaining $\nu$. This step is carried out using Mathematica's \texttt{Eliminate} function on the connection formula. 

Finally, the MST coefficients $a_n^{\nu}$ themselves are recovered from the definitions of the raising and lowering ratios in~\eqref{eq:raising} and~\eqref{eq:lowering} and the fact that conventionally $a_0^{\nu} = 1$. Notably, sometimes $\nu$ is required to an order or two higher than $\tilde{n}_0$ to get the coefficients to $\tilde{n}_0$. This was the case for $s = 0$ and $l = 1$, with $\nu$ required to order 4 to get $a_1^{\nu}$ to order 3. Such issues are peculiarities of the order relationships from~\eqref{eq:raising} and~\eqref{eq:lowering}, and arise due to certain terms in the numerators or denominators of $\alpha_n^{\nu}$, $\beta_n^{\nu}$, and $\gamma_n^{\nu}$ vanishing, changing their order in $\epsilon$. A general way to check the order needed of $\nu$ for determination of the coefficients is not practical to implement, so each use of the algorithm requires a prior consideration of order relationships.

\begin{figure}[H]
    \centering
    \includegraphics[width=1\linewidth]{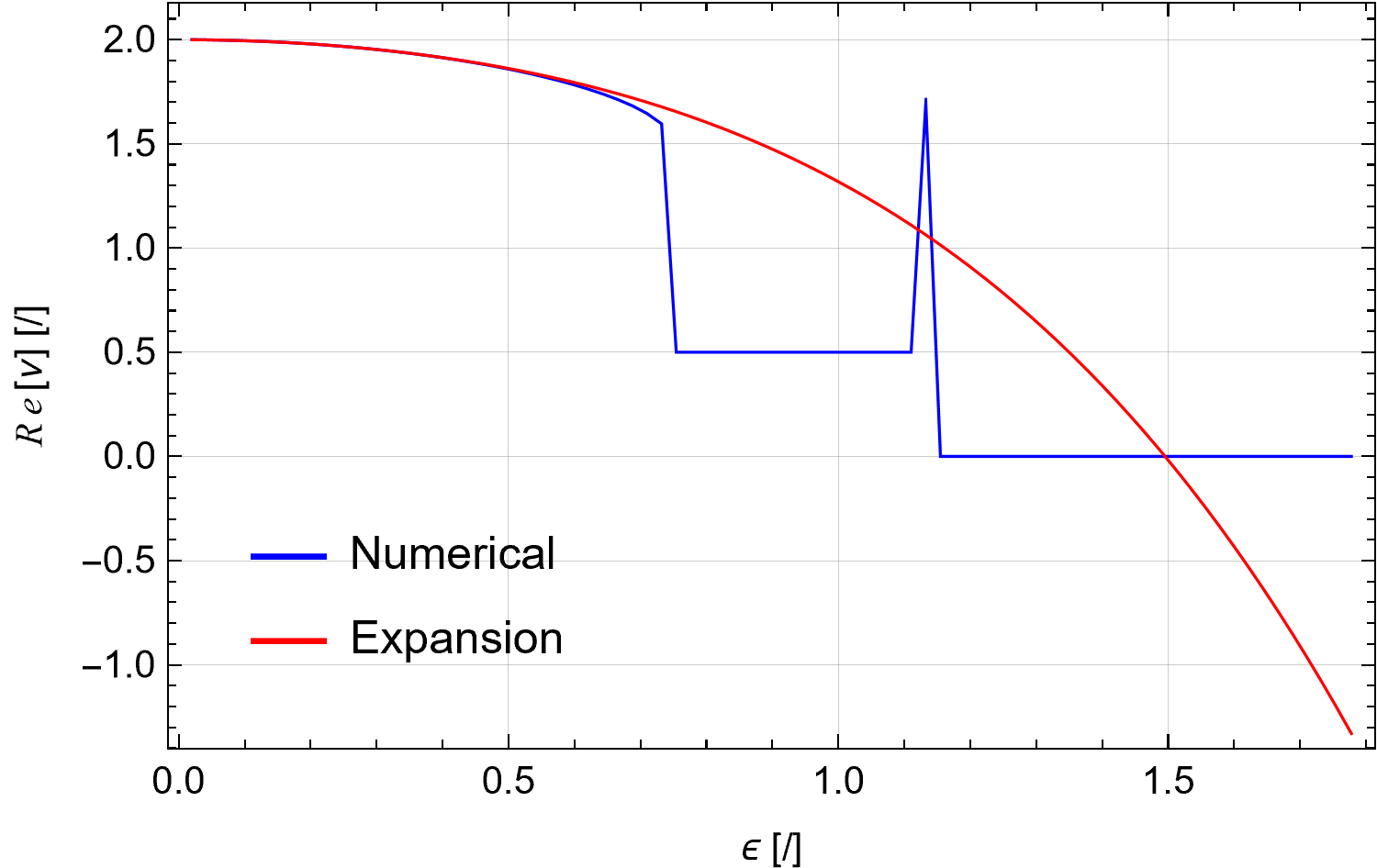}
    \caption{The expansion of $Re[\nu]$ up to fourth order in $\epsilon$ and the numerical result of $Re[\nu]$ for $s = 2,\ l = 2,\ m = 0$. The analytic expansion and numerical result match well until around $\epsilon = 0.7$, which defines the radius of convergence of the analytic 
    expression for the renormalized angular momentum. 
    }
    \label{fig:renorm}
\end{figure}

The obtained expansions of $\nu$ and the coefficients were found to match ones in \cite{Mano:1996vt} for general $l$, as well as for non-general $l$ and $s = 0$ (which changes the order of a few ratios) in \cite{Casals:2016soq}. This signifies the method is working as expected. As an illustration, in Figure \ref{fig:renorm}, the real part of $\nu$ obtained using BHPT toolkit's Teukolsky package was plotted against the analytic expansion up to fourth order in $\epsilon$ for $s = 2,\ l = 2,\ m = 0$, showing an expected trend. As laid out in \cite{Fujita:2005kng}, the discontinuities and jumps of the numerical solution arise due to $\nu$ obtaining a non-negligible imaginary part at appreciable values of $\epsilon$. In an eikonal approximation to the scattering problem, this imaginary part arises due to the shadow of the black hole in the impact parameter space \cite{Ivanov:2025ozg}. The analytical expansion in low $\epsilon$, however, does not exhibit the jumps and has no imaginary part due to the limited radius of convergence.

\subsection{The scattering phase factor}
After obtaining the coefficients to order $\tilde{n_0}$, they can now be used to get the phase factor $\delta_{lm\omega}$ to order $n_0$ by inserting them into equations~\eqref{eq:Binc},~\eqref{eq:Brefl}, and~\eqref{eq:radsol}, carefully making sure the desired order is obtained after expanding at each step. This corresponds to steps 2 and 3 in Figure \ref{fig:flowchart}. To save on computational resources, this step is carried out in three phases in Mathematica. Firstly, $A^{\nu}_{\pm}$ are expanded using~\eqref{eq:Aplus} and~\eqref{eq:Aminus}, and then, along with the exponential prefactors $e^{\pm i(\epsilon \ln \epsilon - \frac{1 - \kappa}{2}\epsilon)}$, their ratio is expanded. This is known as the far zone contribution to the phase factor. Secondly, $K_{\nu}$ and $K_{-\nu - 1}$ are expanded using their formulas from \cite{Bautista:2022wjf}, equation (A.11), by expanding first their prefactors, then the larger, more computationally intensive sum factors. They are then combined to yield the remainder of the formula for $_{-s}B^{(refl)}_{lm\omega} / _{-s}B^{(inc)}_{lm\omega}$. This is known as the near zone contribution to the phase factor. Finally, the far zone and near zone contributions are combined to give the phase factor from~\eqref{eq:fulldelta}.

For $s = 0$, the obtained factor matched the expansion up to the provided order 3 in $\epsilon$ in \cite{Ivanov:2024sds}. More generally than there, the expansions for $s = 0$ with $l = 0$ and $l = 1$ were recovered in terms of $q \neq 0$. The same was done for $s = 1$ with $l = 1$ and $m = 0$, as well as $s = 2$ with $l = 2$, $m = 0$, and $P = -1$. Table \ref{tab:coeff2} shows the newly calculated coefficients for $s = 2$, while in appendix A, Table \ref{tab:coeff1} shows them for $s = 1$, and Table \ref{tab:coeff0} shows them for $s = 0$. In addition, the expansion for general $l$ and $m$ was carried out for $s = 0$. Table \ref{tab:genLs0} in Appendix A shows these results. Though technically possible to obtain too, general $l$ expansions were not carried out for $s = 1$ and $s = 2$ due to a significant computation time needed. We note, similar to \cite{Ivanov:2024sds}, that for the real part of the phase for $s = 0$ it holds
\begin{multline}
    Re[_0\delta_{lm\omega}] = \epsilon \textrm{ln}(2\epsilon) + \sum_{n = 1}^{3} {}_0c_{n, lm\omega}\epsilon^n \\ + \delta_{\left[l0\right]}\epsilon^3\left[\frac{3}{2} - \frac{11\kappa}{12} - \gamma_E - \textrm{ln}(2\kappa \epsilon)\right] + O(\epsilon^4),
\end{multline}
where $\delta_{\left[l0\right]}$ is the Kronecker delta, while for $s = 1, 2$ we have just
\begin{equation}
    Re[_s\delta_{lm\omega}] = \epsilon \textrm{ln}(2\epsilon) + \sum_{n = 1}^{3} {}_sc_{n, lm\omega}\epsilon^n + O(\epsilon^4).
\end{equation}
For the exponentiated imaginary part of the phase factor, which is necessary to determine the transmission factor in~\eqref{eq:trans}, for $s = 0, 1, 2$ we introduce the expansion
\begin{equation}
    e^{-2Im[_s\delta_{lm\omega}]} = 1 + \sum_{n = 1}^3 {}_st_{n,lm\omega}\epsilon^n + O(\epsilon^4).
\end{equation}

\renewcommand{\arraystretch}{1.5}

\begin{table}[!hbt]
    \centering
    \caption{Expansion coefficients $_{s}c_{n,lm\omega}$ and $_{s}t_{n,lm\omega}$ of $Re[\delta_{lm\omega}]$ and $e^{-2Im[\delta_{lm\omega}]}$ for $s = 2, \; l = 2,\; m = 0,\; P = -1$}
    \begin{tabular}{|c||c||c|}
         \hline
         & $_2c_{n,20\omega}$ & $_2t_{n,20\omega}$ \\
         \hline
         \hline
        $n = 1$ & $-\frac{5}{3} + \gamma_{E}$ & 0 \\ 
        \hline
        $n = 2$ & $\frac{107}{420}\pi$ & 0 \\ 
        \hline
        $n = 3$ & $\frac{107}{1260}\pi^2 - \frac{1}{3}\zeta(3) - \frac{1}{24}q^2 - \frac{29}{648}$ & 0 \\ 
        \hline
    \end{tabular}
    \label{tab:coeff2}
\end{table}
\begin{figure}[H]
    \centering
    \includegraphics[width=1\linewidth]{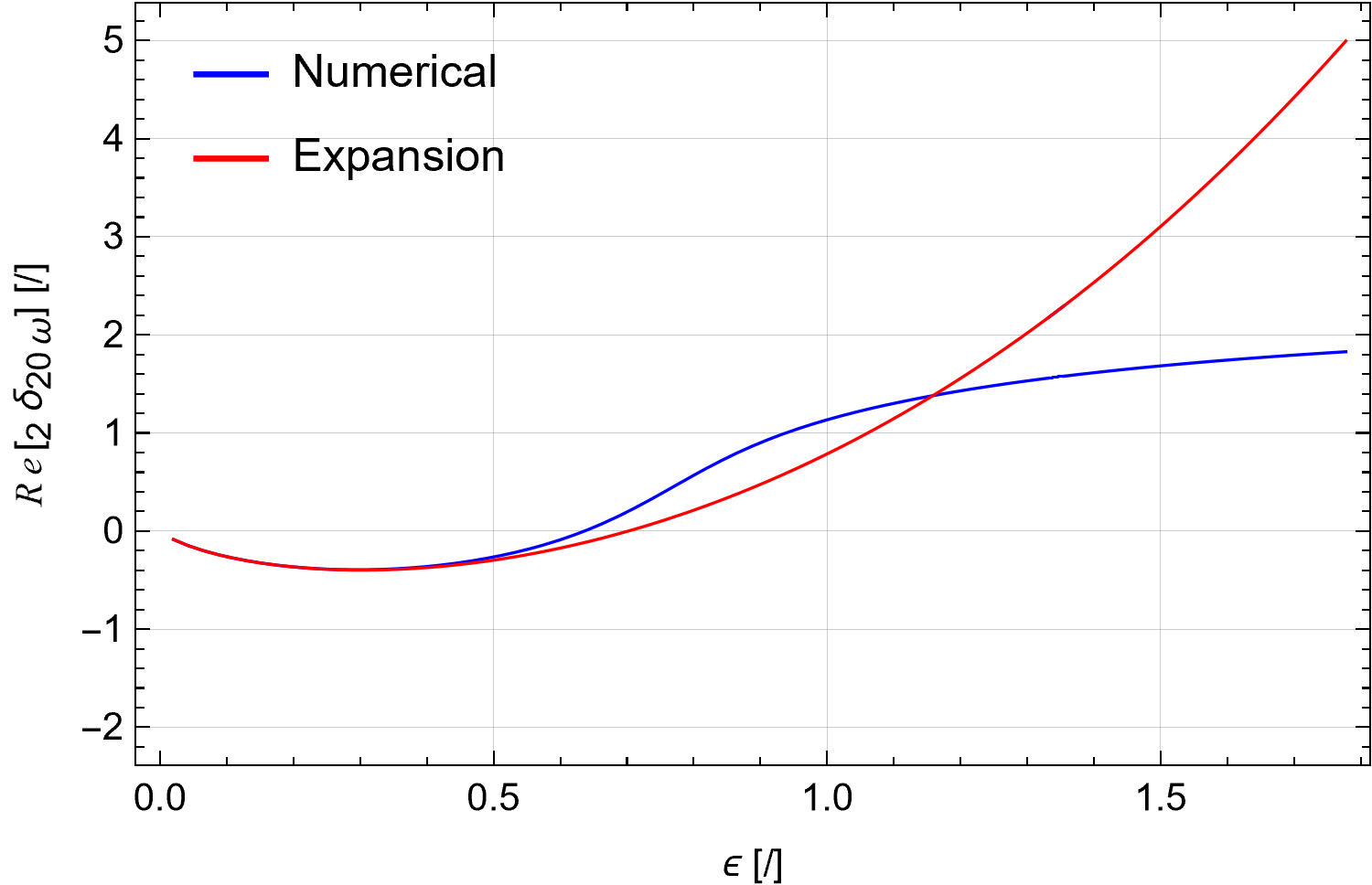}
    \caption{The analytic expansion of $Re\left[_2\delta_{20\omega}\right]$ for $q = 0.5$ and $P = -1$ up to third order in $\epsilon$ and the corresponding numerical result. Once again, agreement is seen until around $\epsilon = 0.75$, where the numerical result exhibits an inflection point, the same value of $\epsilon$ shown to be the convergence radius of the analytic expression for $\nu$ in this case.}
    \label{fig:phaseRe2}
\end{figure}

As a visual comparison, in Figure \ref{fig:phaseRe2} the obtained analytic expansion up to third order in $\epsilon$ of $Re[\delta_{lm\omega}]$ for $s = 2,\; l = 2,\; m = 0$ was compared to a precise numerical MST result obtained using the BHPT toolkit. For completeness, in Appendix B, the comparison process was repeated for $s = 1$ with $l = 1$ and $m = 0$ in Figure \ref{fig:phaseRe1} as well as for $s = 0$ with $l = 0, 1$ and $m = 0$ in Figures \ref{fig:phaseRe00} and \ref{fig:phaseRe01}. All the trends generally showed an expected shape here as well. Finally, also in Appendix B, the numerical result for the mode absorption factor, $e^{-2Im\left[\delta_{lm\omega}\right]}$, was compared with its expansion up to third order for $s = 0$ and $s = 1$ in Figures \ref{fig:eta00} and \ref{fig:eta1}, though for the smaller region $\epsilon \leq 0.5$ as divergence is observed earlier here. The $s = 2$ result was omitted in this case as this is a high order effect.

\section{Conclusions}
\label{sec:concl}

Progress in 
the interpretation of gravitational 
wave data requires 
precision waveform models 
whose development 
benefits from
the scattering amplitude computations. A particularly
important problem in this context
is the scattering of gravitational waves off isolated compact objects,
which allows for a systematic
on-shell matching of tidal 
effects. This problem, 
called the gravitational Raman scattering, 
is especially insightful as 
its amplitude
can be computed 
exactly in general relativity
in the absence of back-reaction.
This a rare example
of an exact highly-nontrivial 
scattering amplitude that 
encodes rich gravitational physics. 

In this paper we develop 
and describe a 
module for both analytic and numerical 
computations of Raman scattering 
amplitudes in general relativity. Our module 
parallels the well known 
black hole perturbation theory toolkit
~\cite{BHPToolkit}, but in addition
to numerical computations of scattering amplitudes, it allows one to extract analytic expressions. 
We reproduce some of the known 
results in the literature, 
and also produce new results
for the scattering phase shifts
and inelasticity parameters
for Kerr black holes
in the post-Minkowskian limit. 
We also explore the issue of convergence of the analytic and numerical expansions of BHPT scattering amplitudes. 

To summarize, we hope 
that our analytic toolkit will be 
a useful addition to 
computational modules
utilized by the black hole perturbation
theory and scattering 
amplitudes communities.

\section*{Acknowledgments}
We thank Zihan Zhou for useful discussions and feedback on the draft. We also thank Scott Hughes for his encouragement of this work.

\onecolumngrid
\appendix
\label{app}

\section{Expansion coefficients for $s = 0$, $s = 1$, and $s = 2$}

\begin{table*}[htb!]
    \centering
    \caption{Expansion coefficients $_sc_{n,lm\omega}$ of $Re[\delta_{lm\omega}]$ for $s = 0$ and general $l, m$.}
    \begin{tabular}{|c||c|}
        \hline
        & $_0c_{n,lm\omega}$ \\
        \hline
        \hline
        $n = 1$ & $-\frac{1}{2} - \psi^{(0)}(1 + l)$\\
        \hline
        $n = 2$ & $\frac{(-11 + 15l + 15l^2)\pi}{4(-1 + 2l)(1 + 2l)(3 + 2l)} - \frac{mq}{2l(1 + l)}$\\
        \hline
        $n = 3$ & $\frac{-11 + 15l + 15l^2}{2(-1 + 2l)(1 + 2l)(3 + 2l)}\psi^{(1)}(1 + l) + \frac{1}{6}\psi^{(2)}(1 + l) + \frac{9(1 + 2l)(-2 + 3l(1 + l)) - 6(-3 + 5l(1 + l))mq\pi - 3(1 + 2l)(l + l^2 - 3m^2)q^2}{12l(1 + l)(-1 + 2l)(1 + 2l)(3 + 2l)}$\\
        \hline
    \end{tabular}
    \label{tab:genLs0}
\end{table*}

\begin{table*}[htb!]
    \centering
    \caption{Expansion coefficients $_sc_{n,lm\omega}$ and $t_{n,lm\omega}$ of $Re[\delta_{lm\omega}]$ and $e^{-2Im[\delta_{lm\omega}]}$ for $s = 0$ and a few specific $l, m$.}
    \begin{tabular}{|c||c|c||c|}
        \hline
        $l$ & $0$ & $1$ & $0$ \\ 
        \hline
        $m$ & \multicolumn{2}{c||}{$0$} & $0$ \\ 
        \hline
        \hline
          & \multicolumn{2}{c||}{$_0c_{n,lm\omega}$} & $_0t_{n,lm\omega}$ \\
          \hline
        \hline
        $n = 1$ & $-\frac{1}{2} + \gamma_{E}$ & $-\frac{3}{2} + \gamma_{E}$ & 0 \\ 
        \hline
        $n = 2$ & $\frac{11\pi}{12}$ & $\frac{19\pi}{60}$ & $-\kappa - 1$ \\ 
        \hline
        $n = 3$ & $\frac{11\pi^2}{36} - \frac{1}{3}\zeta(3) - \frac{1}{12} + \frac{q^2}{12}$ & $\frac{19\pi^2}{180} - \frac{1}{3}\zeta(3) - \frac{1}{20}q^2$ & $-\kappa\pi - \pi$ \\ \hline
    \end{tabular}
    \label{tab:coeff0}
\end{table*}
% Rewrite paragraph 1 more clearly
% In paragraph 2 mention that expanding the fraction causes the divergence.

\begin{table*}[htb!]
    \centering
    \caption{Expansion coefficients $_sc_{n,lm\omega}$ and $t_{n,lm\omega}$ of $Re[\delta_{lm\omega}]$ and $e^{-2Im[\delta_{lm\omega}]}$ for $s = 1, \; l = 1,\; m = 0$.}
    \begin{tabular}{|c||c||c|}
         \hline
         & $_1c_{n,10\omega}$ & $_1t_{n,10\omega}$ \\
         \hline
         \hline
        $n = 1$ & $-\frac{5}{2} + \gamma_{E}$ & 0 \\ 
        \hline
        $n = 2$ & $\frac{47}{120}\pi$ & $\frac{q^8}{8}$ \\ 
        \hline
        $n = 3$ & $\frac{47}{360}\pi^2 - \frac{1}{3}\zeta(3) - \frac{1}{8}q^2 + \frac{7}{96}$ & 0 \\ 
        \hline
    \end{tabular}
    \label{tab:coeff1}
\end{table*}

\newpage

\section{Comparisons between numerical results and the expansions for $s = 0$ and $s = 1$.}

\begin{figure}[htb!]
    \centering
    \includegraphics[width=0.75\linewidth]{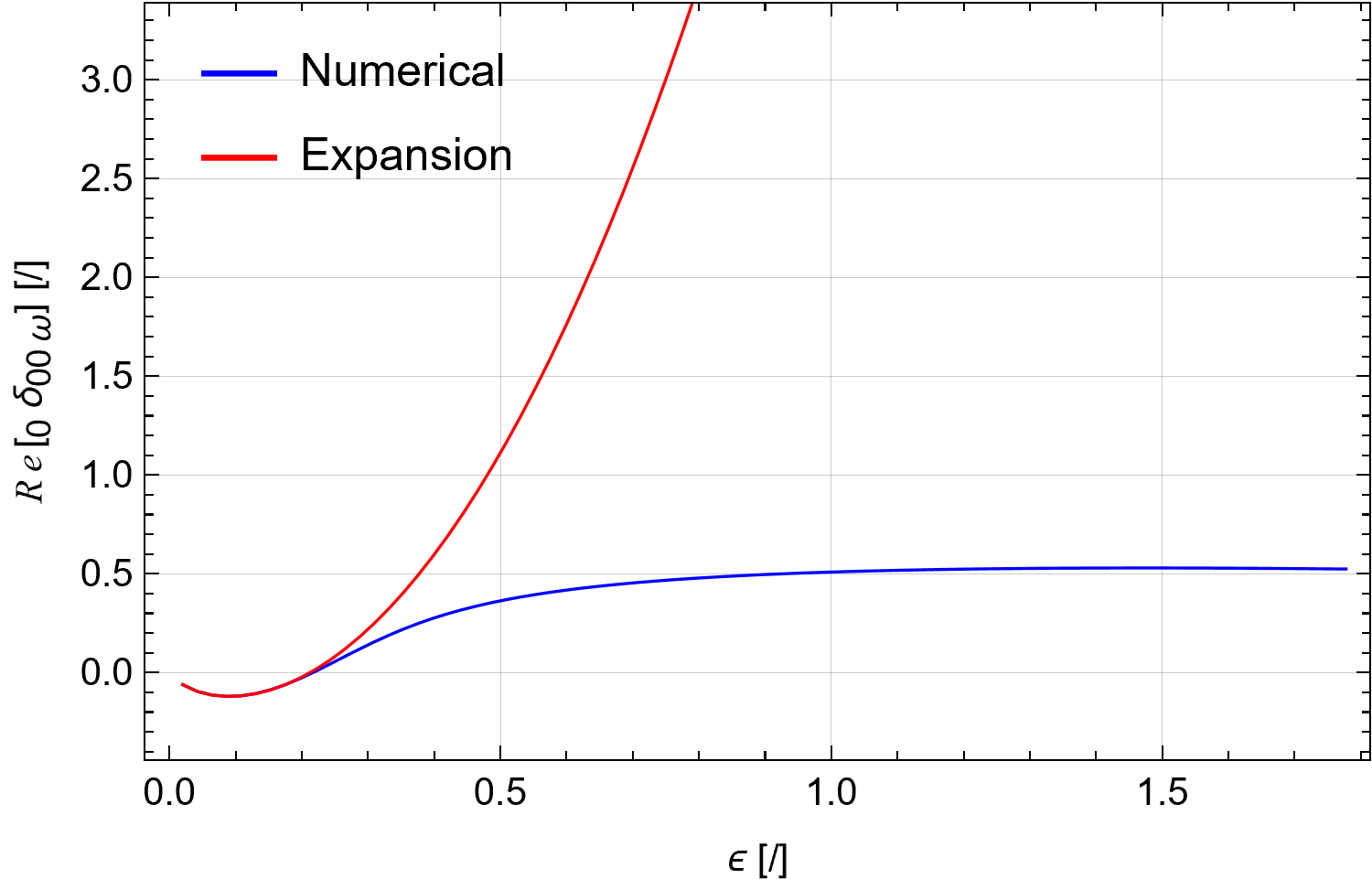}
    \caption{The analytic expansion of $Re\left[_0\delta_{00\omega}\right]$ for $q = 0.5$ up to third order in $\epsilon$ and the corresponding numerical result.}
    \label{fig:phaseRe00}
\end{figure}

\begin{figure}[htb!]
    \centering
    \includegraphics[width=0.75\linewidth]{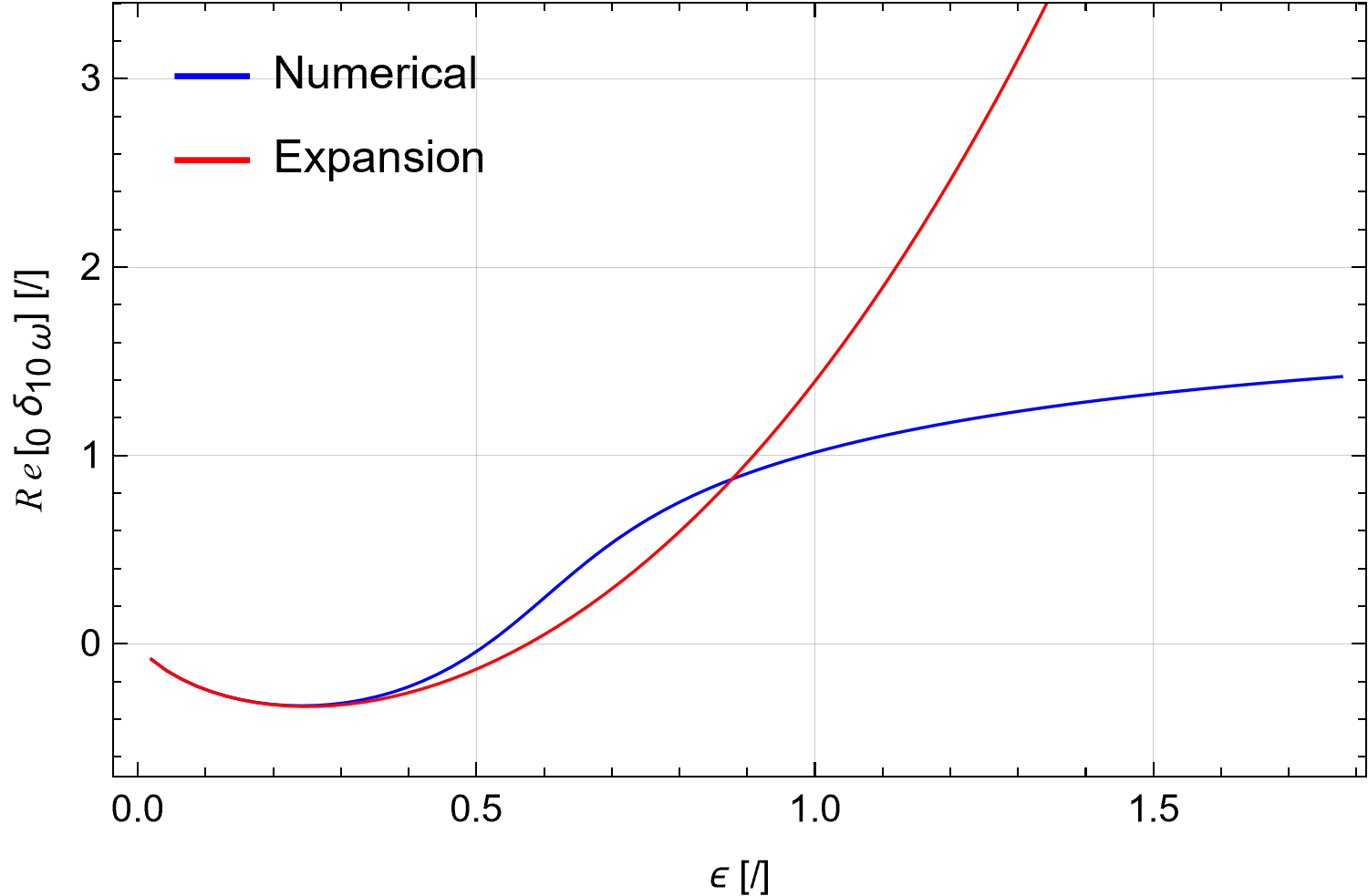}
    \caption{The analytic expansion of $Re\left[_0\delta_{01\omega}\right]$ for $q = 0.5$ up to third order in $\epsilon$ and the corresponding numerical result.}
    \label{fig:phaseRe01}
\end{figure}

\begin{figure}[htb!]
    \centering
    \includegraphics[width=0.75\linewidth]{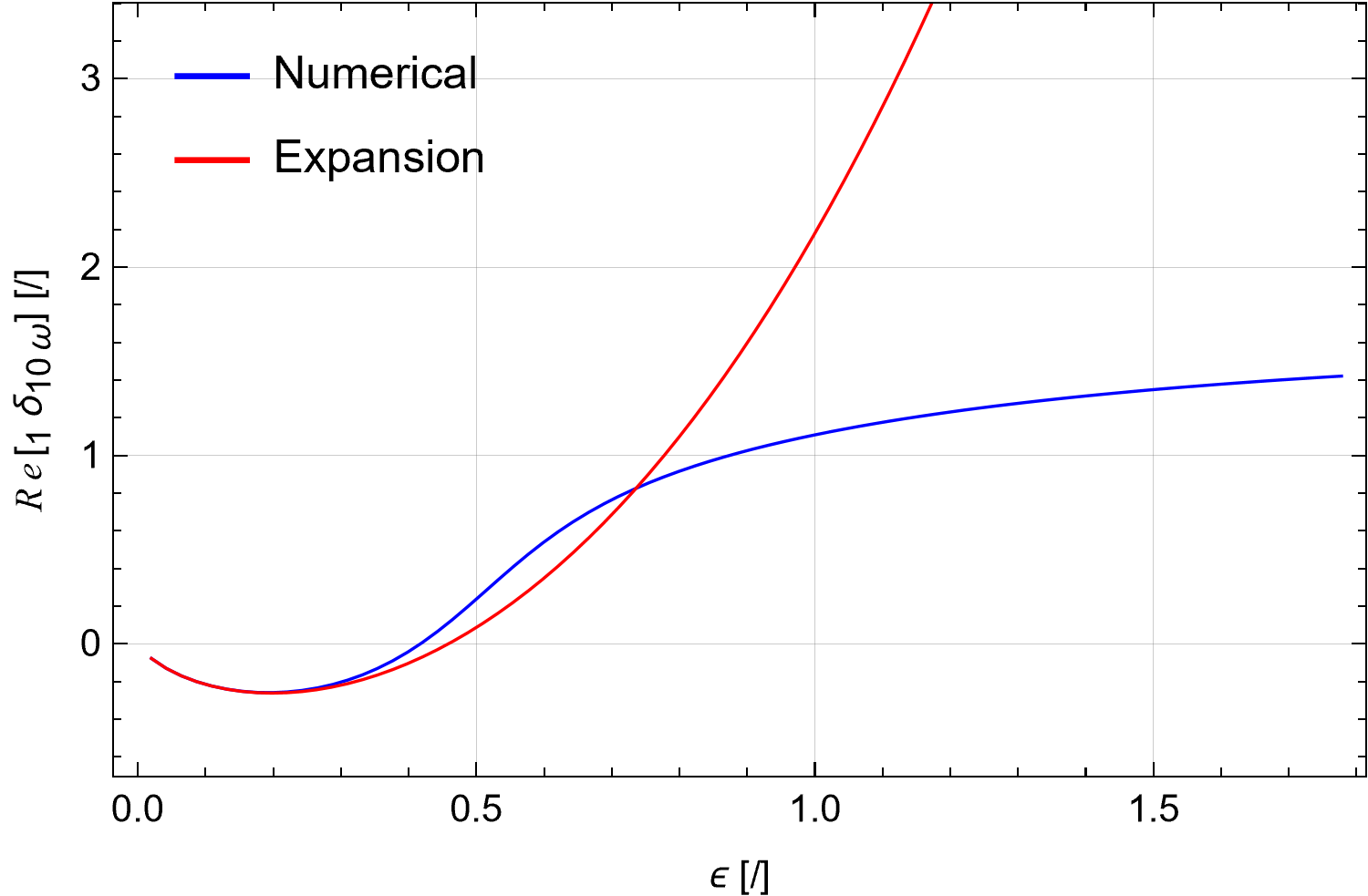}
    \caption{The analytic expansion of $Re\left[_1\delta_{10\omega}\right]$ for $q = 0.5$ up to third order in $\epsilon$ and the corresponding numerical result.}
    \label{fig:phaseRe1}
\end{figure}

\begin{figure}[htb!]
    \centering
    \includegraphics[width=0.75\linewidth]{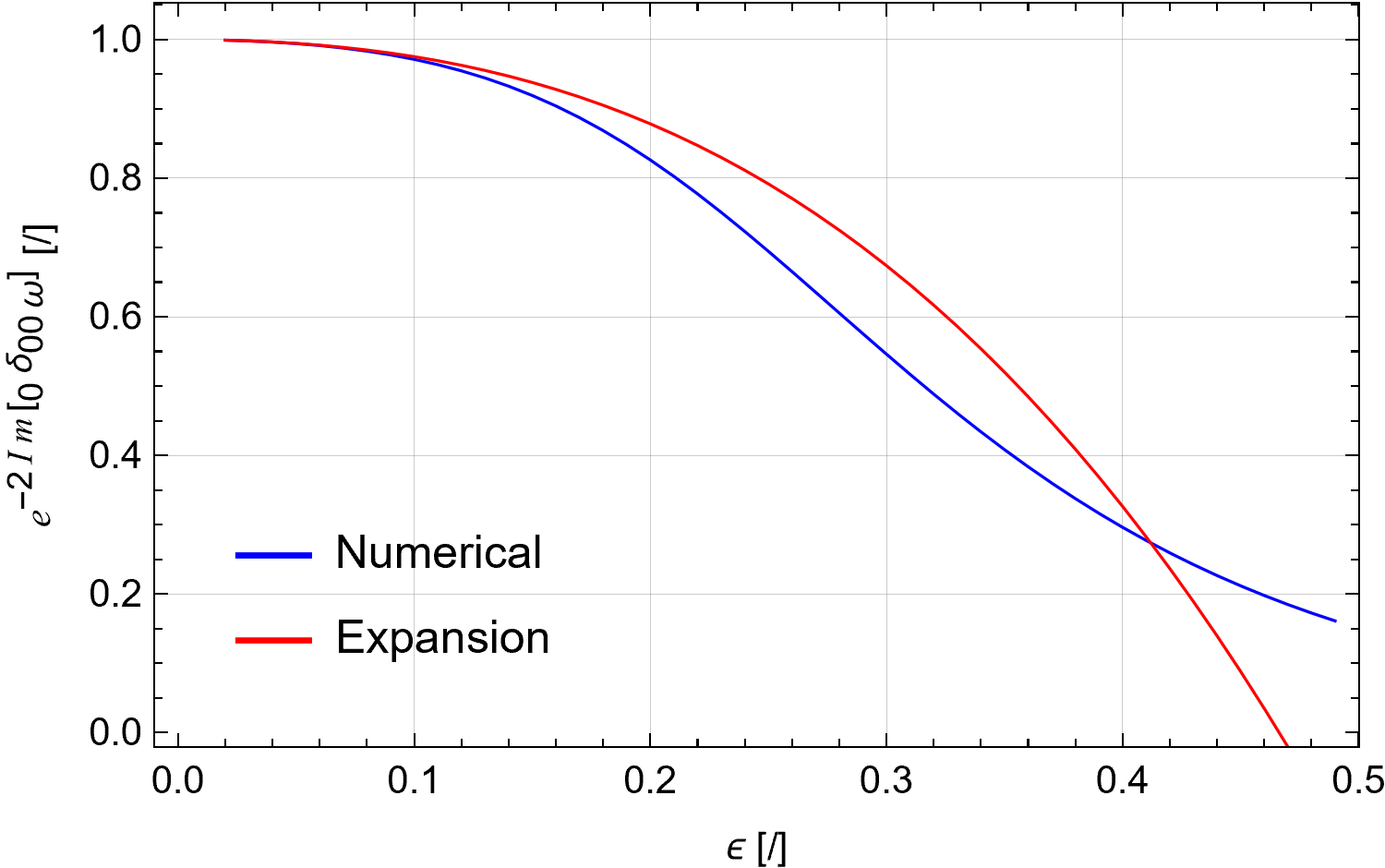}
    \caption{The analytic expansion of $e^{-2Im\left[_0\delta_{00\omega}\right]}$ for $q = 0.5$ up to third order in $\epsilon$ and the corresponding numerical result.}
    \label{fig:eta00}
\end{figure}

\begin{figure}[htb!]
    \centering
    \includegraphics[width=0.75\linewidth]{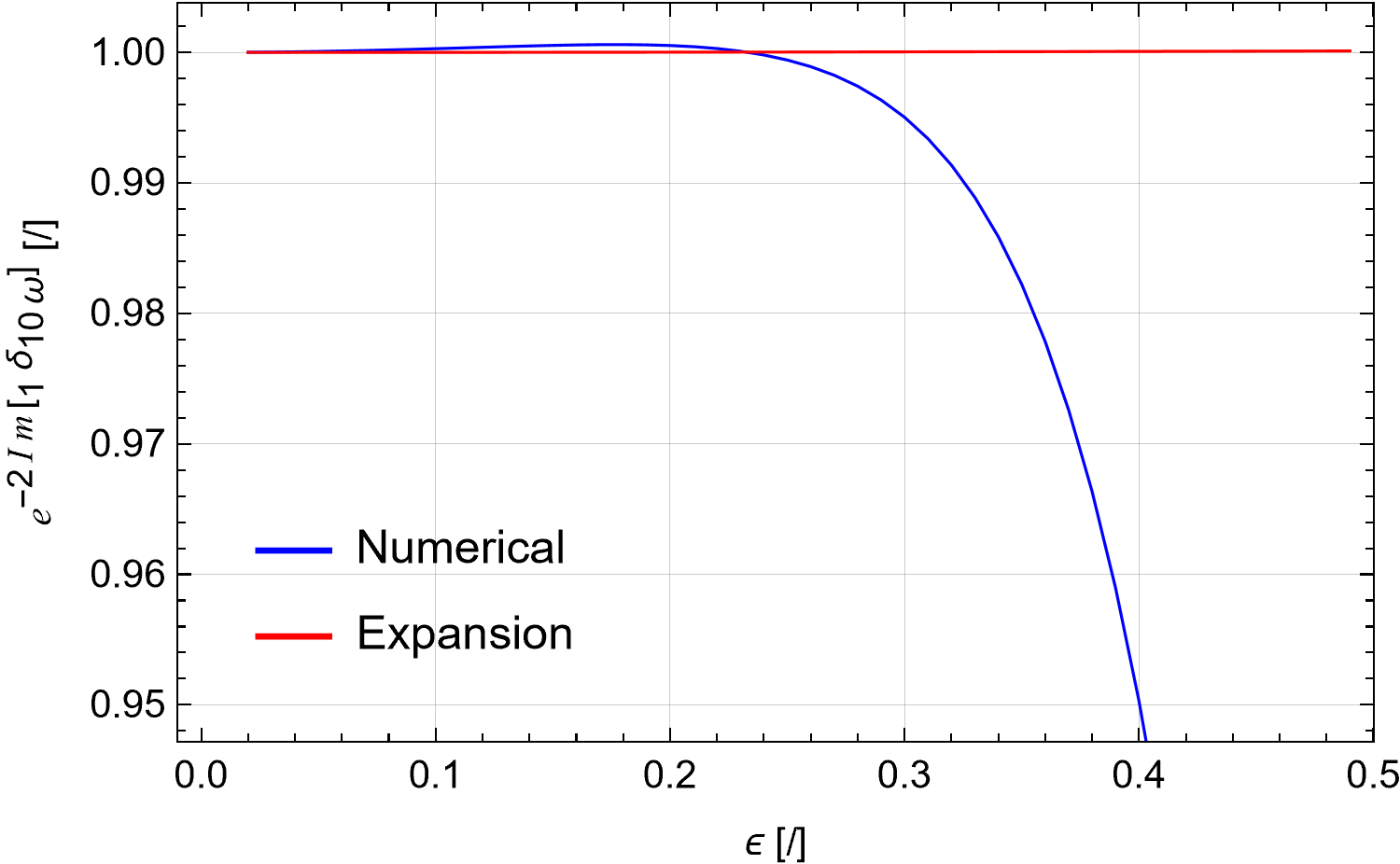}
    \caption{The analytic expansion of $e^{-2Im\left[_1\delta_{10\omega}\right]}$ for $q = 0.5$ up to third order in $\epsilon$ and the corresponding numerical result. Note the expansion unexpectedly hovers around 1. This is due to the low expansion order in $\epsilon$.}
    \label{fig:eta1}
\end{figure}

% \nocite{*}

\clearpage
\bibliography{apssamp.bib}% Produces the bibliography via BibTeX.

\end{document}